\documentclass[twocolumn,twoside,slac_two]{revtex4}
\usepackage{graphicx}
\usepackage{fancyhdr}
\begin{document}

\title{Pathfinder flight of the Polarized Gamma-ray Observer (PoGOLite) in~2013}

%

\author{Takafumi Kawano, On behalf of the PoGOLite Collaboration}
\affiliation{High Energy \& Optical/Infrared Astrophysics Laboratory, Hiroshima University, Higashi-Hiroshima, Hiroshima, 739-8526}
%
	
\begin{abstract}
The Polarized Gamma-ray Observer (PoGOLite) is a balloon-borne instrument that can measure polarization in the energy range 25--240~keV. The instrument adopts an array of well-type "phoswich" detectors in order to suppress backgrounds. Based on the anisotropy of Compton scattering angles resulting from polarized gamma-rays, the polarization of the observed source can be reconstructed. During July 12-26 of 2013, a successful near-circumpolar pathfinder flight was conducted from Esrange, Sweden, to Norilsk, Russia. During this two-week flight, several observations of the Crab were conducted. Here, we present the PoGOLite instrument and summarize the 2013 flight.
￼{\it tkawano@hep01.hepl.hiroshima-u.ac.jp}.

\end{abstract}

\maketitle

\thispagestyle{fancy}


\section{Measuring polarization}
 
Astrophysical phenomena can be observed with electromagnetic radiation by imaging, 
spectroscopy, timing analysis, and polarimetry.
Since X-ray observations began 50 years ago, many observatories have provided data for imaging, 
spectroscopy, and timing analyses.
On the other hand, polarization measurements of X-rays and gamma-rays have been technically difficult 
and only a few sensitive observations have been performed.
Polarized X-rays and gamma-rays are expected to be emitted from a wide variety of astronomical sources, including pulsars,
X-ray binary systems, strongly magnetized neutron stars, collimated outflows from active galactic nuclei and gamma-ray bursts.
Therefore, polarimetric studies of these sources are expected to provide important new insight into the physics of such highly energetic objects.
In particular, it is important to understand the acceleration site and magnetic field structure of pulsars and their surrounding wide nebulae by identifying the emission mechanisms with polarimetry.

\section{Target object: the Crab nebula}
\subsection{Outline}
The Crab nebula is a remnant of the historical supernova in 1054 A.D., located around 2~kpc from the Earth.
This celestial object is named "Crab nebula" after the characteristic filament structure in optical wavelengths, 
and it has been studied intensively in all wavelengths from radio to gamma-rays since the early days of astronomy. 
The Crab consists of a pulsar, a synchrotron nebula and a bright expanding shell of thermal gas. 
We can also see a highly collimated bipolar outflow (jet), which is aligned to the spin axis of the pulsar,
as well as a circumstellar torus visible in X-rays. The high-energy emission is brightest near the center of the nebula.

 The Crab pulsar is considered as a neutron star with a radius of 10 km, a mass of 1.4 $M_{\odot}$, a rotation period \mbox{P = 33~ms}, 
\mbox{$\dot{P}$ = 4.21~$\times$~10$^{-13}$}, magnetic field \mbox{B~$\approx$~10$^{12}$~G}, 
and spin-down luminosity \mbox{$L_{s}$ $\approx$ 5 $\times$ 10$^{38}$~erg/s}. 
The strong magnetic field and short rotation period produce a relativistic outflow of electron-positron pairs which is called the pulsar wind. 
This ultrarelativistic outflow is confined by the thermal ejecta. 
The inner ring of the nebula, reported to be at a distance of about 3~$\times$~10$^{17}$~cm from the pulsar, corresponds to a termination shock, created when the pulsar wind interacts with the surrounding synchrotron nebula. 
It is considered that the pulsar wind and possibly other particles are accelerated to energies 
as high as $\sim$10$^{16}$~eV at the termination shock.
High energy charged particles interact with the magnetic field in the nebula ($\sim$a~few~mG), 
and emit synchrotron radiation.
The X-ray emission becomes softer toward the outer region owing to adiabatic and radiative losses.
At the edge of the nebula, there are only low-energy radio-emitting electrons.
The spectrum of X-rays and gamma-rays below 1~GeV for the Crab nebula is well described by synchrotron emission, and inverse Compton scattering dominates above 1~GeV~\cite{crab-outline}.

\subsection{Previous polarization measurements of the Crab nebula}

\paragraph{Measurement by OSO-8}
The OSO-8 satellite carried an X-ray polarimeter consisting of a panel of mosaic graphite crystals, which were utilized for Bragg reflection~\cite{crab-oso8}.

\begin{figure}[htbp]	
 \begin{center}
 \includegraphics[width=75mm]{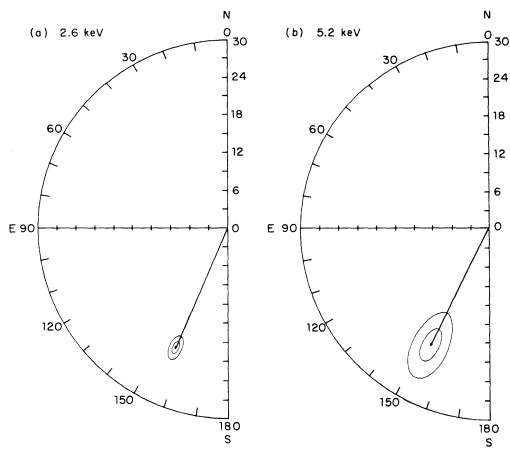}
 \end{center}
 \caption{The polarization vectors for the Crab nebula at (a) 2.6~keV and (b) 5.2~keV. 
 Surrounding regions (in order of increasing size) correspond to the 67\% and 99\% confidence contours~\cite{crab-oso8}.}
 \label{fig:crab-oso8}
\end{figure}

The polarization fraction of the Crab nebula observed by OSO-8 was (19.19~$\pm$~0.97)\% 
with (156.36~$\pm$~1.44)$^{\circ}$ polarization angle at 2.6~keV, 
and (19.50~$\pm$~2.77)\% polarization fraction with (152.59~$\pm$~4.04)$^{\circ}$ polarization angle at 5.2~keV, 
where the errors correspond to 67\% confidence contours. 
These results are in agreement with optical polarization measurements (FIG. \ref{fig:crab-oso8})~\cite{crab-oso8}.
 
\paragraph{Observation by INTEGRAL/SPI}
The SPI (spectrometer onboard INTEGRAL; INTErnational Gamma-Ray Astrophysics Laboratory) has a capability for polarization measurements using Compton scattering~\cite{integral-modulation}. 

\begin{figure}[htbp]
\begin{minipage}{0.5\textwidth}
\begin{center}
\makeatother
\includegraphics[width=75mm]{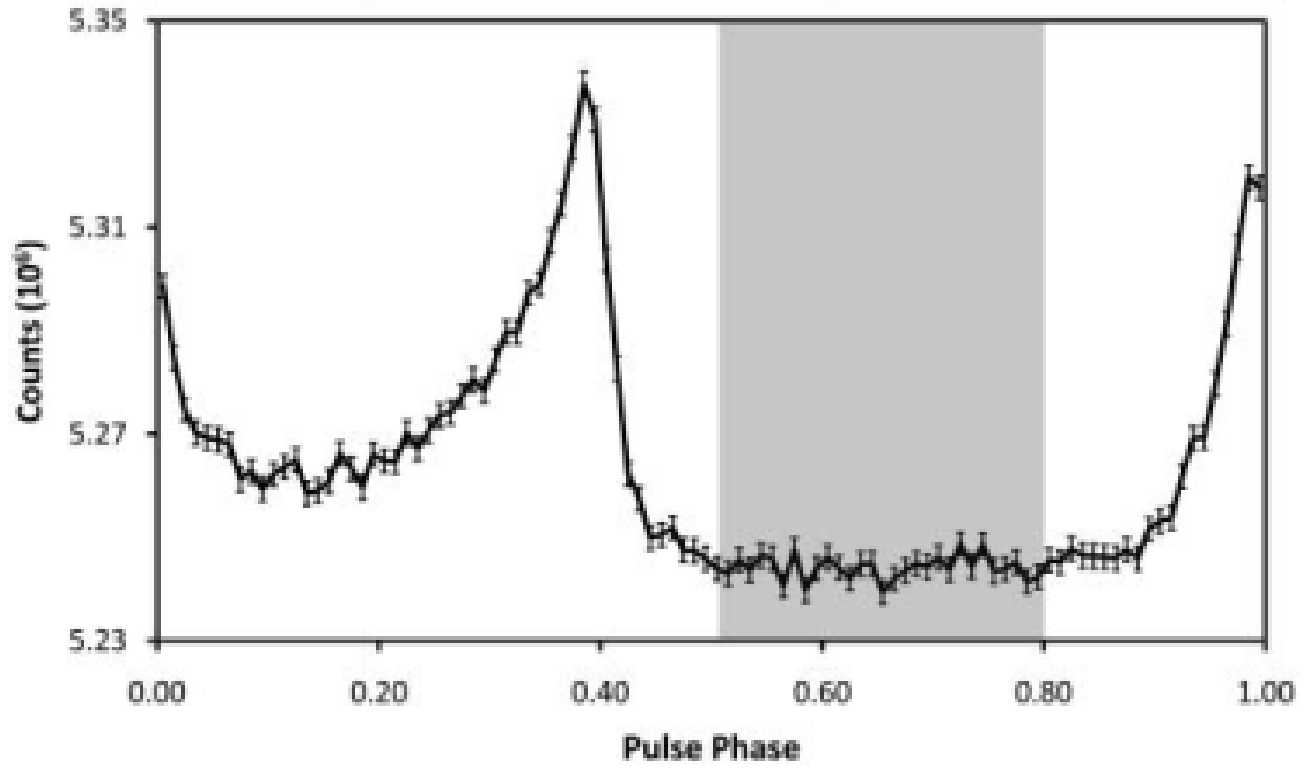}
\end{center}
\end{minipage}
\quad
\begin{minipage}{0.5\textwidth}
\begin{center}
\makeatletter
 \includegraphics[width=70mm]{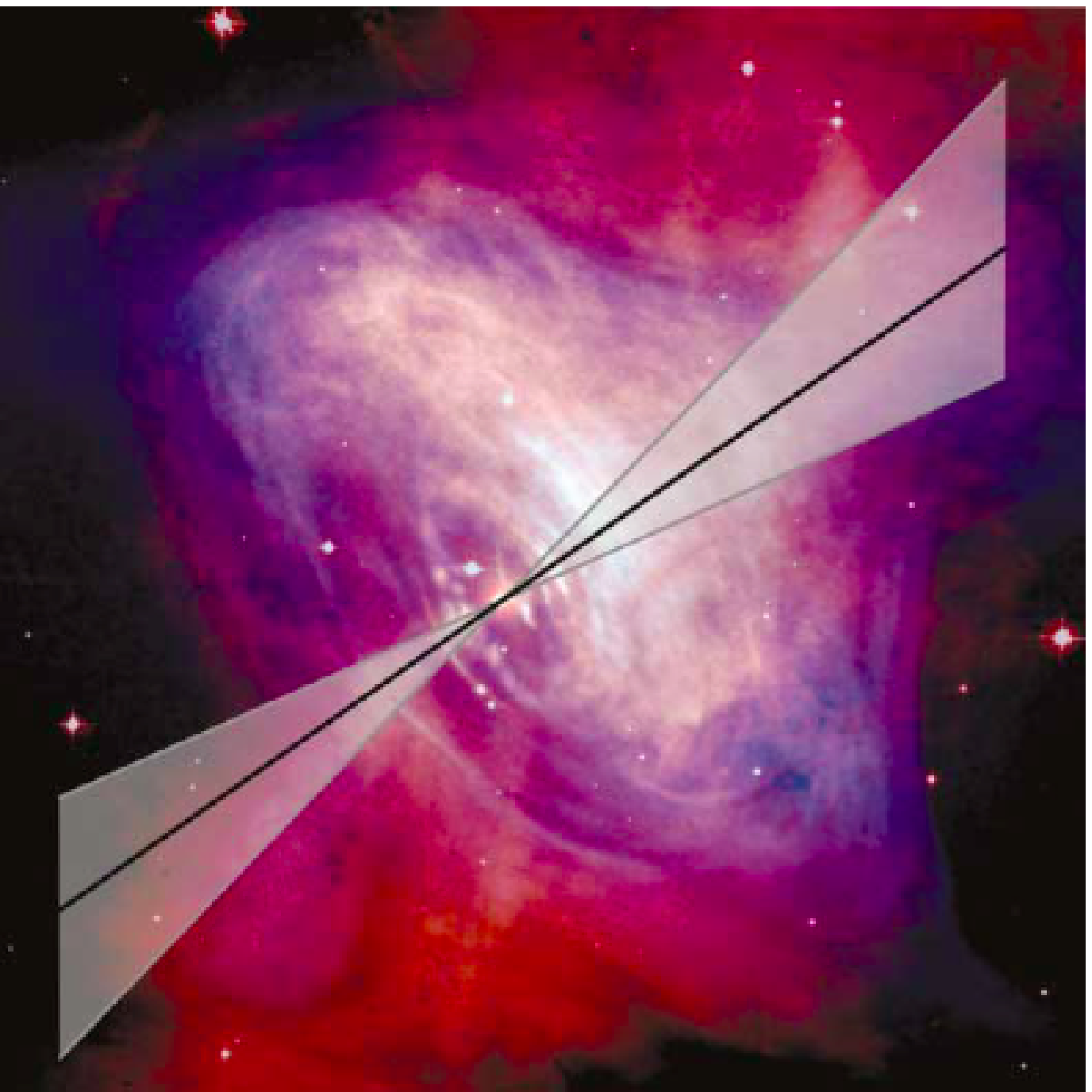}
\end{center}
\end{minipage}
 \caption{Top: The light curve of the Crab pulsar. There are two pulse phases (0.88$<$$\phi$$<$0.14 and 0.25$<$$\phi$$<$0.52), off-pulse phase (0.52$<$$\phi$$<$0.88),
 and last phase called as "Bridge" phase (0.14$<$$\phi$$<$0.25).
 The data for polarization analysis for INTEGRAL/SPI is selected from within the phase interval from 0.5 to 0.8 of the pulsar period (shaded area) in 100--1000~keV.
Bottom: Composite image of the Crab, blue: Chandra X-ray image, 
red: Hubble Space Telescope optical image, 
and the gamma-ray polarization vector is superimposed (gray area).
The direction of the polarization vector is along the jet axis.~\cite{crab-SPI}}
 \label{fig:crab-spi}
\end{figure}

A polarization analysis of the Crab nebula was performed with data recorded from February 2003 to April 2006, 
and only events during the off-pulse fraction of the pulsar cycle were included (FIG. \ref{fig:crab-spi} top).

A polarization fraction of (46~$\pm$~10)\% was observed, with the polarization angle (123~$\pm$~11)$^{\circ}$, 
which is closely aligned with the pulsar spin axis ((124~$\pm$~0.1)$^{\circ}$) (FIG. \ref{fig:crab-spi} bottom), 
but the errors are dominated by systematic effects.
The observed alignment of the polarization axis with the jet axis suggests an orthogonal magnetic field configuration towards the jet axis if the soft gamma-ray emission is caused by synchrotron emission.
The observed polarization fraction is quite high, but less than the maximum limit for synchrotron radiation, 
$\sim$75\%~\cite{crab-SPI}\cite{maxime}.

\paragraph{Observation by INTEGRAL/IBIS} 
The IBIS (imager on INTEGRAL) is also using Compton scattering for polarimetry~\cite{crab-ibis}.
Also this instrument has been used to study the polarization of the Crab.

\begin{figure}[htbp]	
 \begin{center}
 \includegraphics[width=80mm]{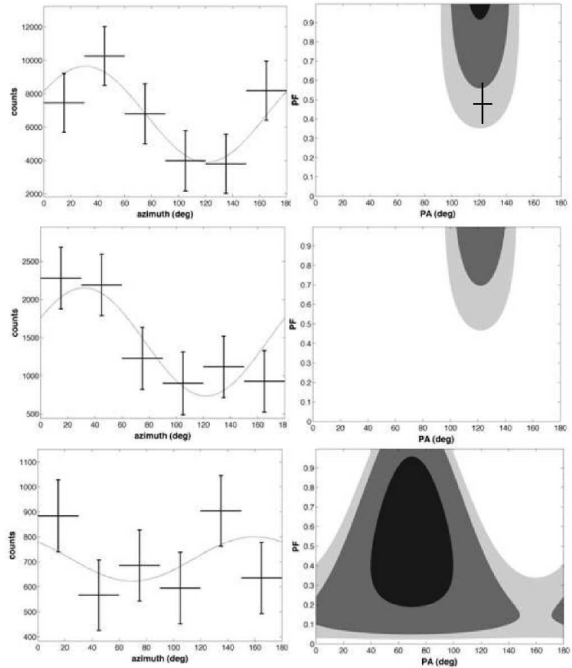}
 \end{center}
 \caption{
The polarization of the Crab nebula observed by INTEGRAL/IBIS. 
The polarization angle and polarization fraction are measured for the Crab data between 200--800~keV, 
in the off-pulse (top), off-pulse and bridge (middle), and two-peak (bottom) phase intervals. 
The error bars for the profile are at 1$\sigma$. The 68\%, 95\%, and 99\% confidence regions are shaded from dark to light gray. 
The SPI result~\cite{crab-SPI} is indicated in the top figure by a cross~\cite{crab-ibis}.}
   \label{fig:ibis-polari}
\end{figure}

The results are shown with respect to the pulsar phase (FIG. \ref{fig:crab-spi} top). 
For the "off-pulse" phase and "off-pulse and bridge" phase, 
the polarization fraction is reported to be quite high ($>$72\%), 
with the polarization angle aligned along the jet axis (FIG. \ref{fig:ibis-polari}).
This result suggests that the off-pulse polarized emission recorded above 200~keV can come from a striped wind, jets, and/or equatorial wind near the bright knot.
The magnetohydrodynamics models predict that the polarization is strongest at the pulsar, in the knot, 
and along the jets, and it should be mostly parallel to the rotation axis~\cite{crab-ibis}.

\section{The PoGOLite balloon-borne instrument}
\subsection{Overview}
\label{overview}
\begin{figure}[htbp]
 \begin{center}
 \includegraphics[width=80mm]{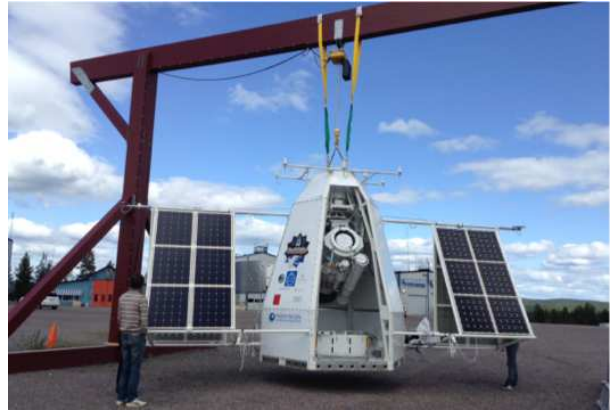}
 \end{center}
 \caption{Overview of the PoGOLite payload. 
 The height is $\sim$4~m and the weight is $\sim$2~tonnes. 
 A 1.1~million~cubic~meter helium-filled balloon is used for lifting the payload.}
  \label{fig:pogo}
\end{figure}

PoGOLite (Polarized Gamma-ray Observer Light-weight version) is a balloon-borne Compton polarimeter (FIG. \ref{fig:pogo}), measuring the polarization of hard X-rays/soft gamma-rays 
from celestial objects in the energy range 25--240~keV~\cite{mozsi}.
Polarization in the 25--100~keV energy band has not been observed previously.

Observational targets for PoGOLite include the Crab and Cygnus X-1, 
and the instrument is able to detect 10\% liner polarization from 
the Crab nebula for 15-hour exposure time with the 99\% confidence, 
in a signal-to-background scenario of 1:1.
The detector is optimized for point sources, with a narrow field of view of 2.4$^{\circ}$~$\times$~2.6$^{\circ}$,
and the required pointing precision is 0.1$^{\circ}$.
Since radiation from sources follows an inverse power-law and photons are additionally absorbed in the atmosphere, 
a high float altitude of the instrument ($\sim$40~km) and sensitivity extending as low as possible are crucial.
The collaboration is international, involving institutes and universities from 
Japan\footnote{Hiroshima University, Tokyo Institute of Technology, Yamagata University, Japan Aerospace Exploration Agency (JAXA).}, 
Sweden\footnote{Royal Institute of Technology (KTH), Stockholm University (SU).}, 
and the United States\footnote{Stanford Linear Accelerator Center (SLAC), Kavli Institute for Particle Astrophysics and Cosmology (KIPAC), University of Hawaii.}.

\begin{figure}[htbp]
 \begin{center}
 \includegraphics[width=75mm]{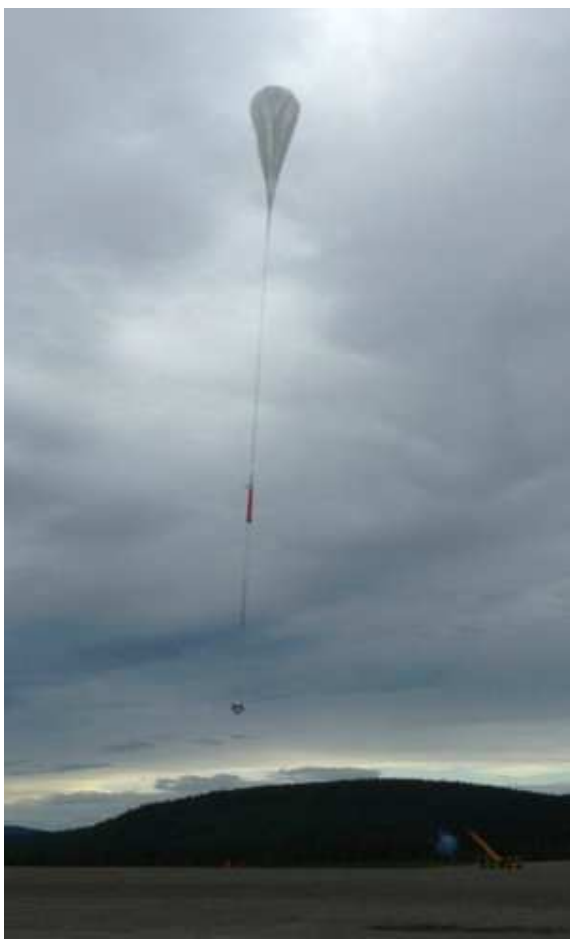}
 \end{center}
 \caption{The launch of PoGOLite from Esrange on July 12th 2013, 0818 UT. The distance from the top of the balloon to the gondola is $\sim$300~m.}
 \label{fig:pogo-pic}
\end{figure}

The full-size version of PoGOLite consists of 217 units. 
It is intended to be able to measure as low as 10\% 
polarization from a 200 mCrab source in a six-hour flight~\cite{kamae}.
The 61-unit "Pathfinder" version of PoGOLite 
has been prepared for launch from Esrange in northen Sweden in 2011, 2012 and 2013.
In this paper, we simply refer to the "PoGOLite Pathfinder" as "PoGOLite".
On July 6th, 2011 (UTC), 
the payload was launched for a flight with a foreseen landing in Canada (duration $\sim$5~days).
However, there was a leak of helium from the balloon, 
and the gondola was returned to ground after $\sim$5~hours.
A second launch was foreseen in July 2012, but had to be cancelled due to unfavorable weather conditions.
PoGOLite was successfully launched from Esrange, Sweden, 
on July 12th at 0818 UT in 2013 (FIG. \ref{fig:pogo-pic}). 
A circumpolar flight was possible thanks to permission received from Russian authorities. 
The flight ended on July 26th when the gondola touched down close to the Siberian city of Norilsk 
($\sim$3000~km to the East of Moscow) at 0015 UT.

\subsection{Detector configuration}
\begin{figure}[htbp]
\begin{minipage}{0.5\textwidth}
\begin{center}
\makeatletter
\includegraphics[width=55mm]{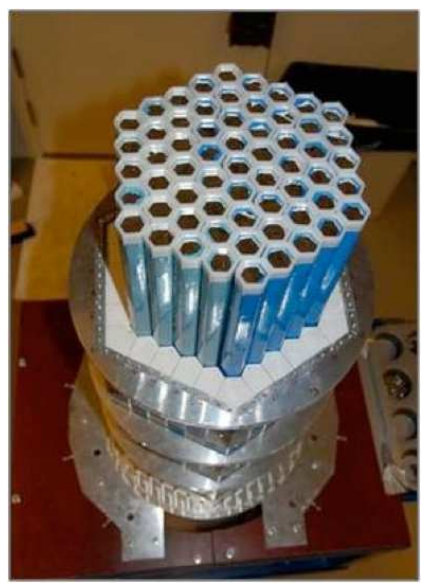}
\end{center}
\end{minipage}
\quad
\begin{minipage}{0.5\textwidth}
\begin{center}
\makeatother
\includegraphics[width=75mm]{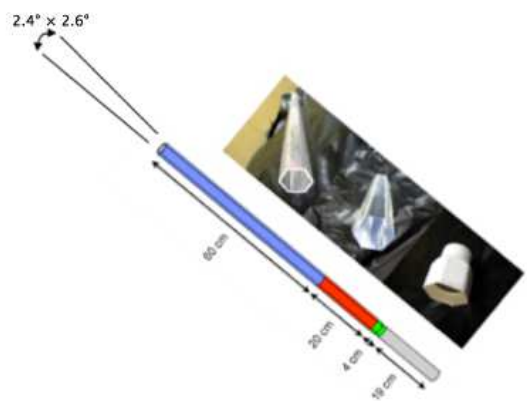}
\end{center}
\end{minipage}
\caption{Top: The main detector consists of 61 PDCs and 30 SASs with BGO. 
Bottom: One of the PDCs, which is consisting of a hollow "slow" plastic scintillator (60 cm), 
a solid "fast" plastic scintillator (20~cm), a BGO crystal (4~cm) and photomultiplier tube (from Hamamatsu Photonics, 19~cm). }
\label{fig:PDC}
\end{figure}

To suppress the high rate of background events at float altitude ($\sim$40~km), 
we adopted 
an array of 61 well-type phoswich detector cells (PDCs) surrounded by a segmented BGO (bismuth germanate oxide, Bi$_{4}$Ge$_{3}$O$_{12}$) anticoincidence shield comprising 30~units. 
A LiCAF (LiCAl$_{6}$) neutron-sensitive scintillator~\cite{hirotaka} is also included (FIG. \ref{fig:PDC} top).
Each PDC consists of three active components: a hollow "slow" plastic scintillator (60~cm), 
a solid "fast" plastic scintillator (20~cm), and a BGO crystal (4~cm), 
read out by a photomultiplier tube (from Hamamatsu Photonics, 19~cm) 
(FIG. \ref{fig:PDC} bottom).
The LiCAF scintillator is made for neutron detection since neutrons are expected to dominate the background.
This detector is sandwiched between two BGO elements for rejecting gamma-rays, allowing neutron interactions to be distinguished.

\begin{figure}[htbp]
 \begin{center}
 \includegraphics[width=70mm]{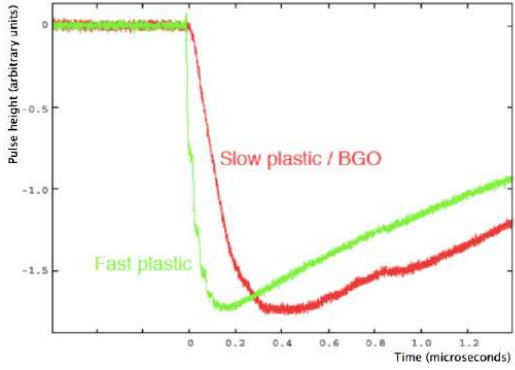}
 \end{center}
 \caption{
Examples of characteristic waveforms (shown with a negative polarity). 
The rise time is shorter for a signal from the fast plastic scintillator than for one from the slow scintillator or BGO crystal.}
 \label{fig:wave}
\end{figure}

\begin{figure}[htbp]
 \begin{center}
 \includegraphics[width=60mm]{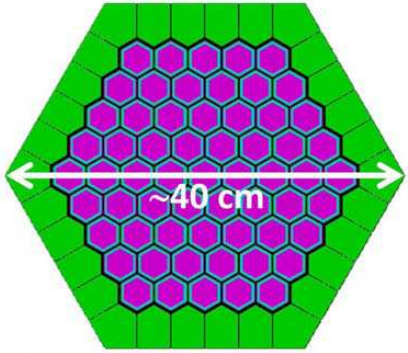}
 \end{center}
 \caption{Top view of a 61-unit detector array. The PDCs (purple) are surrounded by a segmented side anti-coincidence shield (green).}
 \label{fig:TopView}
\end{figure}

The photomultiplier tube waveforms are sampled at a 37.5 MHz rate and digitized with to 12 bit accuracy. The "fast" scintillator, "slow" scintillator and BGO crystal have different decay times, resulting in different pulse shapes for waveforms originating from these components (FIG. \ref{fig:wave}). 
By identifying these differences, 
we can determine in which component an interaction has taken place, allowing background events to be discarded.
The PDC units are hexagonal so they can be tightly packed in a honey-comb structure, 
while surrounding SAS segments have two different pentagonal shapes to fit closely around detector array (FIG. \ref{fig:TopView}).

\subsection{Polarimeter design}
An indicator of performance of X-ray and gamma-ray polarimeters,
 which is called as MDP (Minimum Detectable Polarization;
  degree distinguishable from statistical fluctuation with 3$\sigma$), is written in equation~(1).
\[
MDP = \frac{4.29}{M\times R_{S}}\sqrt{\frac{R_{S}+R_{B}}{T}} \quad (1)
\]
 
where $M$ is the modulation factor (depends on the instrument geometry and the spectrum of the incident photon flux), $R_{S}$ is the signal rate, $R_{B}$ is the background rate and $T$ is the exposure time.
This represents the minimum polarization fraction measurable by the instrument for the given confidence level~\cite{mdp}.
A large $M$, a large $R_{S}$, a small $R_{B}$, and a large $T$ are needed to achieve a low MDP, i.e. good sensitivity to polarization.
For a given source, a large $R_{S}$ corresponds to a large effective area of the instrument.

For PoGOLite, the effective area is $\sim$22 cm$^{2}$, 
with a reasonable modulation factor (M $\sim$26\% at 50~keV)~\cite{maxime_pogo}.

\section{Measuring polarization}
\begin{figure}[htbp]
 \begin{center}
 \includegraphics[width=50mm]{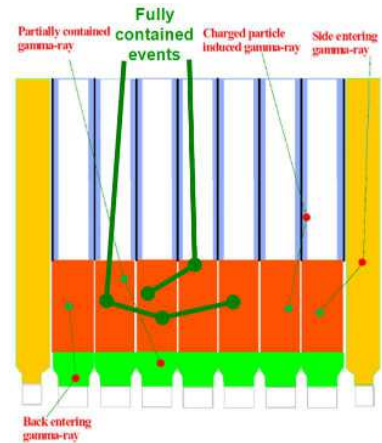}
 \end{center}
\caption{The detection method of polarization.
Cross section of the detector, tracking both of scattering and absorption position of X-rays.
Some examples of background events have also been indicated.}
\label{fig:HowToDetect}
\end{figure}

The PoGOLite instrument is using Compton scattering for polarimetry.
The procedure is illustrated below (FIG. \ref{fig:HowToDetect}):
\begin{itemize}
 \item  Polarized gamma-rays undergo Compton scattering in a hexagonal array of plastic scintillators.
 \item  Polarized photons tend to scatter perpendicularly to the polarization direction, following equation~(2).
 \item  Observed azimuthal scattering angles are modulated by polarization.
\end{itemize}

\[
\frac{d\sigma}{d\Omega} = \frac{1}{2}r_{e}^{2}\frac{k^{2}}{k_{0}^{2}}\biggl(\frac{k}{k_{0}}+\frac{k_{0}}{k}  - 2\ sin^{2}\theta cos^{2} \phi \biggr)
\quad (2)
\]

where $k_{0}$ and $k$ are the momenta of the incident and scattered photon, respectively, 
r$_{e}$ is the classical electron radius and $\theta$ and $\phi$ correspond to the polar and azimuthal scattering angles. 
Angle $\phi$ is defined relative to the polarization direction of the incident photon, 
resulting in a polarization-dependence for the scattering process.

Tracking individual photons through coincident detection of Compton scattering and photoelectric absorption allows the azimuthal scattering angles to be reconstructed. 
Since photons scatter preferentially perpendicular to the polarization direction, the resulting distribution of scattering angles will be anisotropic (modulated) for a polarized flux of photons.

\section{Flight trajectory \& Results}
\begin{figure}[htbp]
 \begin{center}
 \includegraphics[width=80mm]{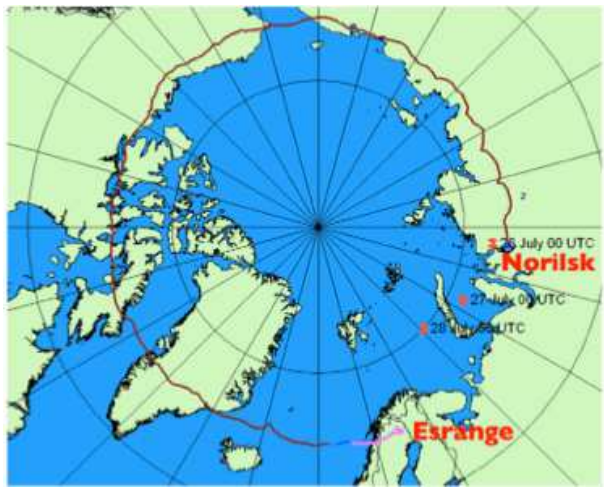}
 \end{center}
 \caption{The trajectory of PoGOLite flight in 2013. 
PoGOLite was launched from Esrange, Sweden at 0818 UT on July 12th in 2013, 
and landed Norilsk, Russia at 0015 UT on July 26th.
Courtesy of SSC Esrange.}
 \label{fig:map}
\end{figure}

As mentioned in section \ref{overview}, 
during July 12-26 of 2013, a successful near-circumpolar pathfinder flight was conducted from Esrange, 
Sweden, to Norilsk, Russia (FIG. \ref{fig:map}). 
During the daytime, the balloon has higher altitude because of heating of the helium inside the balloon. 
Conversely, the balloon has lower altitude during the night, resulting in a diurnal variation of the pressure.

The attitude control system performance has been evaluated from Crab measurements, 
and the observed performance was found to be an order of magnitude better than the design requirement of 0.1$^{\circ}$.

We obtained the pulse-folded light curve of the Crab pulsar.
The data includes two- and three-hit events in the energy range 20~keV -- 110~keV, and we can clearly see the Crab pulsation from the light-curve.
This shows that X-ray photons from the Crab pulsar are indeed detected by the polarimeter, confirming that the instrument and attitude control system are working as intended.

\section{Summary \& Outlook}

During July 2013, the PoGOLite Pathfinder made a circumpolar flight ($\sim$13.5~days) from Esrange. This flight was possible thanks to permission from the Russian government and help from Russian colleagues.
We have confirmed that the polarimeter detected the Crab pulsation. 
Crab polarization results are in preparation.
We have a plan to improve the polarimeter design based on the experience from the 2013 flight, to reject background more efficiently.
Reflight of PoGOLite is proposed for the summer of 2016.

\bigskip 
%

\bigskip 

\end{document}